\documentclass[a4paper,11pt]{article}
\usepackage[english]{babel}
\usepackage{jheppub}
\usepackage{subfig}
\usepackage[T1]{fontenc} 
\usepackage{graphicx}
\usepackage{epsfig}
\usepackage{axodraw}
\usepackage{amsmath}
\usepackage{amssymb}
\usepackage{float}
\usepackage{placeins}

\allowdisplaybreaks[4]

\title{Two-Loop Master Integrals for the Planar QCD Massive Corrections to Di-photon and Di-jet Hadro-production}

\author[a,b]{Matteo Becchetti}
\author[a,b]{Roberto Bonciani}

\affiliation[a]{Sapienza - Universit\`a di Roma, Dipartimento di Fisica, Piazzale Aldo Moro 5, 00185, Rome, Italy}

\affiliation[b]{INFN Sezione di Roma, Piazzale Aldo Moro 2, 00185, Rome, Italy}

\emailAdd{matteo.becchetti@roma1.infn.it}
\emailAdd{roberto.bonciani@roma1.infn.it}

\abstract{We present the analytic calculation of the Master Integrals necessary to compute the planar
  massive QCD corrections to Di-photon (and Di-jet) production at hadron colliders. The masters are evaluated
  by means of the differential equations method and expressed in terms of multiple polylogarithms and one- or two-fold integrals of polylogarithms and irrational functions, up to transcendentality four.}

\newcommand{\be}{\begin{equation}}
\newcommand{\ee}{\end{equation}}
\newcommand{\nn}{\nonumber}
\newcommand{\bea}{\begin{eqnarray}}
\newcommand{\eea}{\end{eqnarray}}
\newcommand{\bfig}{\begin{figure}}
\newcommand{\efig}{\end{figure}}
\newcommand{\bc}{\begin{center}}
\newcommand{\ec}{\end{center}}

\begin{document}


%
%
%
%

\maketitle
\flushbottom

\section{Introduction}

In this paper, we consider the calculation of the two-loop Master Integrals (MIs) needed for the evaluation of the massive NNLO QCD planar corrections to the hadro-production of a photon pair (di-photon production). The same two-loop masters enter the evaluation of the massive NNLO QCD planar corrections to the hadro-production of two jets (a pair of gluons, $q \bar{q}$ or $q(\bar{q})g$ pairs, di-jet production). For massive QCD corrections we mean QCD corrections that contain a loop of heavy-quarks.

The two processes are very relevant for the physics programme at the LHC.
Di-photon production, due to an experimentally clean final state, provides an important test of the Standard Model (SM)
and constitutes an irreducible background for a Higgs, produced in gluon fusion, that decays into two photons.
The current theoretical description of the di-photon production includes the NNLO QCD corrections \cite{Catani:2011qz} due to
massless states, and a part of the N$^3$LO massless corrections in the gluon-gluon channel \cite{Bern:2002jx}, while the massive corrections (at the two-loop level) are not included\footnote{The massive corrections are included, at the NNLO, in the gluon-gluon channel via the interference of two one-loop diagrams, the so-called {\it box contribution} \cite{Dicus:1987fk}.}.
Di-jet production is a dominant process at the LHC. It enters the determination of the strong coupling constant, it is
sensible to the value of parton distribution functions and it is important for searches of new physics, beyond the SM.
Very recently the  NNLO QCD corrections in the purely gluonic channel were computed \cite{Currie:2013dwa}. In
\cite{Currie:2017eqf}, a phenomenological study of the dijet production, doubly differential in the mass of the dijet system and in the rapidity difference, was presented. The study involves NNLO QCD corrections to all the partonic channels, leading in the number of colors. These perturbative corrections include massless states and the massive corrections (at the two-loop level) are not considered in the analysis.

If we expand the di-photon production cross section in powers of the couplings, the fine structure constant $\alpha$ and the strong coupling constant $\alpha_S$, the order $\alpha^2 \alpha_S^2$ is the first perturbative order at which QCD massive corrections appear. In the $q \bar{q}$ channel, this gives a genuine two-loop correction, interference between the two-loop and the tree-level matrix elements. In the $gg$ channel, since the photons do not couple to the gluons,
the order $\alpha^2 \alpha_S^2$ contains only the interference between one-loop diagrams, in which a heavy-quark box mediates the coupling between gluons and photons. Two-loop corrections appear, in this channel, at the following perturbative order, $\alpha^2 \alpha_S^3$ and arise from the interference of two-loop with one-loop diagrams.

For the di-jet cross section, the leading order (LO), $\mathcal{O}(\alpha_S^2)$, is present in both the partonic channels. Therefore, NNLO QCD corrections contain genuine two-loop corrections initiated by a $q\bar{q}$, $gg$ or $q(\bar{q})g$ pairs. In the case of the partonic process $q\bar{q} \to q\bar{q}$ the massive corrections involve a closed heavy-quark loop as correction to the gluon propagator.
For the partonic processes $q\bar{q} \to gg$, $gg \to q\bar{q}$ and $q(\bar{q})g \to q(\bar{q})g$, the massive corrections involve instead actual two-loop massive box diagrams, as the ones used for the di-photon production. The channel $gg \to gg$ needs additional massive box diagrams, with respect to the ones presented in this paper; some of them are known already in the literature \cite{Caron-Huot:2014lda}.

The di-photon production partonic cross section has a simple color structure. At the NNLO, it is proportional to $C_FC_A$, where $C_F$ is the Casimir of the fundamental representation of $SU(N_c)$ and $C_A$ is the Casimir of the adjoint representation. This implies that both planar and crossed Feynman diagrams contribute to the sole gauge-invariant color coefficient.
The di-jet production partonic cross section has, instead, a complicated structure in terms of color coefficients. In this case,
the planar diagrams contribute to all the color coefficients, while the crossed diagrams do not enter the leading color one.
This means that, the computation of the planar diagrams alone allows, nevertheless, to give physical predictions in the leading color approximation, while for the di-photon case, this is not possible.

We approach the calculation using Feynman diagrams. We construct the interference between the two-loop and the tree-level
amplitudes; this is expressed as a combination of dimensionally regularized scalar integrals. These integrals are reduced
to a set of 37 MIs using the computer programs\footnote{Other public programs are available
  for the reduction to the MIs \cite{Anastasiou:2004vj,Lee:2012cn,Lee:2013mka,Maierhoefer:2017hyi}.}
\texttt{FIRE} \cite{Smirnov:2008iw,Smirnov:2013dia,Smirnov:2014hma} and
\texttt{Reduze 2} \cite{Studerus:2009ye,vonManteuffel:2012np}, that implement integration-by-parts identities
\cite{Tkachov:1981wb,Chetyrkin:1981qh,Laporta:2001dd} and Lorentz-invariance identities \cite{Gehrmann:1999as}.

The MIs are computed using the differential equations method \cite{Kotikov:1990kg,Remiddi:1997ny,Gehrmann:1999as,Argeri:2007up,Henn:2014qga}. The system of differential equations obeyed by the MIs is cast in canonical form \cite{Henn:2013pwa} (see also
\cite{Argeri:2014qva,DiVita:2014pza,Henn:2013nsa,Gehrmann:2014bfa,Lee:2014ioa,Adams:2017tga,Schneider:2016szq,Meyer:2016slj,Georgoudis:2016wff,Gituliar:2017vzm}). The solution is expressed in terms of
Chen's iterated integrals, represented whenever possible in terms of Goncharov's polylogarithms (GPLs) \cite{Goncharov:polylog,Goncharov2007,Remiddi:1999ew}. The part at transcendentality four of seven four-point functions and one part at transcendentality three are
given in terms of single and double parametric integrations. The reason is that it was not possible to find a change of kinematic
variables able to ``linearize'' the whole set of square roots that belong to the original alphabet.

The analytic results presented in this article are collected in ancillary files which we upload 
with the arXiv submission\footnote{The results expressed in terms of GPLs are provided in text files in GiNaC format, while the single and double parametric integrations are provided in ``.mx'' format, readable by Mathematica 11 \cite{Mathematica}}.

The article is structured as follows. In Section \ref{not}, we give our notations and conventions. In Section \ref{diff}, we discuss the system of differential equations and the canonical form of the set of MIs. Moreover, we present the alphabet that enters the solutions of the differential equations and the transformation of variables that linearizes all the square roots except one. In Section \ref{mis}, we present our basis and the transformation matrix between the MIs in canonical and in ``pre-canonical'' form. Finally, in Section \ref{concl} we conclude. In appendix \ref{appendixa}, we discuss the prescription for the linearization of the square roots and in Appendix \ref{appendixb} we give the routing of the MIs in pre-canonical form.

\section{Notations \label{not}}

We consider the basic processes\footnote{The processes $gg \to \gamma \gamma, (gg,q\bar{q})$, involve additional topologies and additional MIs with respect to the ones we are going to present in this paper.}
$q \bar{q} \to \gamma \gamma,(gg,q\bar{q})$ and the crossed $q(\bar{q})g \to q(\bar{q})g$, in which the initial partons have momenta $p_1$ and $p_2$ and the final photons or partons have momenta $p_3$ and $p_4$. The external particles are on their mass-shell $p_i^2=0$. 

We introduce the Mandelstam variables
\be
s = (p_1+p_2)^2 \, , \quad t = (p_1-p_3)^2 \, , \quad u = (p_1-p_4)^2 \, ,
\ee
such that $s+t+u=0$. Since we consider $2\to 2$ scattering processes with massless external particles, the physical region is defined through the following relations
\be
s > 0 \, , \quad t = - \frac{s}{2} ( 1-\cos{\theta} ) < 0 \, , \quad -s<t<0 \, ,
\ee
where $\theta$, $0<\theta<\pi$, is the scattering angle in the partonic center of mass frame.

For later convenience we define the following dimensionless ratios
\be
\label{2.4a}
u = - \frac{s}{4m_t^2} \, , \quad v = - \frac{t}{4m_t^2} \, ,
\ee
where $m_t$ is the mass of the heavy quark that runs into the loops.

The NNLO QCD planar corrections to the partonic processes listed above can be calculated reducing to the MIs two topologies,
shown in Fig.~\ref{fig1} and defined by the integrals
\be
\int {\mathcal D}^dk_1 {\mathcal D}^dk_2 \frac{D_8^{a_8}D_9^{a_9}}{D_1^{a_1}D_2^{a_2}D_3^{a_3}D_4^{a_4}D_5^{a_5}D_6^{a_6}D_7^{a_7}} \, ,
\label{2.4}
\ee
where $a_i$ are positive integers, the $D_i$, $i=1,...,9$, are the denominators involved, $d$ is the dimension of the space-time,
$\epsilon = \frac{4-d}{2}$, $\gamma_E=0.5772..$ is the Euler-Mascheroni constant and the normalization is such that
\be
   {\mathcal D}^dk_i = \frac{d^d k_i}{i \pi^{\frac{d}{2}}} e^{\epsilon\gamma_E} \left( \frac{m_t^2}{\mu^2} \right) ^{\epsilon} \, .
\ee
The routings that we used for the two topologies are the following:
\bea
\mbox{Topology A} & := & \Bigl\{ -k_1^2, -(p_1-k_1)^2, -(k_1+p_2)^2, m_t^2-k_2^2, m_t^2-(-k_1-k_2+p_1)^2, \nn\\
&& \hspace*{3mm} m_t^2-(k_1+k_2+p_2)^2, m_t^2-(-k_1-k_2+p_1-p_3)^2, -(k_1+p_3)^2, \nn\\
&& \hspace*{3mm} -(k_1+k_2)^2 \Bigr\} \, , \\
\mbox{Topology B} & := & \Bigl\{-k_1^2,-(p_1-k_1)^2,-(k_1+p_2)^2,m_t^2-k_2^2,m_t^2-(-k_1-k_2+p_1-p_3)^2, \nn\\
&& \hspace*{3mm} m_t^2-(k_1+k_2+p_2)^2, -(-k_1+p_1-p_3)^2, (k_2+p_3)^2, (k_1+k_2)^2 \Bigr\} \, .
\eea
The momenta $k_1$ and $k_2$ are the loop momenta.

The number of MIs for the two topologies is 32 for Topology A and 25 for Topology B.
However, some of them appear in both topologies. Therefore, the total number of independent MIs
is 37. Among them, 12 are new and presented in this paper for the first time.
\bfig
\bc
\[ 
\vcenter{
\hbox{
  \begin{picture}(0,0)(0,0)
\SetScale{1.0}
  \SetWidth{1}
\Line(25,-30)(40,-30)
\Line(25,30)(40,30)
\Line(-25,30)(-40,30)
\Line(-25,-30)(-40,-30)
\Line(-25,30)(-25,-30)
\Line(-25,-30)(25,-30)
\Line(25,30)(-25,30)
\Line(-25,-30)(-25,30)
%
%
  \SetWidth{4}
\Line(0,-30)(0,30)
\Line(0,-30)(25,-30)
\Line(25,-30)(25,30)
\Line(25,30)(0,30)
%
%
%
%
\Text(0,-45)[c]{\footnotesize{(A)}}
%
\end{picture}}
}
\hspace*{5cm}
\vcenter{
\hbox{
  \begin{picture}(0,0)(0,0)
\SetScale{1.0}
  \SetWidth{1}
\Line(25,-30)(40,-30)
\Line(25,30)(40,30)
\Line(-25,30)(-40,30)
\Line(-25,-30)(-40,-30)
\Line(-25,30)(-25,-30)
\Line(-25,-30)(25,-30)
\Line(25,30)(-25,30)
\Line(-25,-30)(-25,30)
\Line(0,-30)(25,-30)
\Line(25,-30)(25,30)
\Line(25,30)(0,30)
%
%
  \SetWidth{4}
\Line(0,-30)(25,0)
\Line(25,0)(25,-30)
\Line(0,-30)(25,-30)
%
%
%
%
\Text(0,-45)[c]{\footnotesize{(B)}}
%
\end{picture}}
}
\]
\vspace*{7mm}
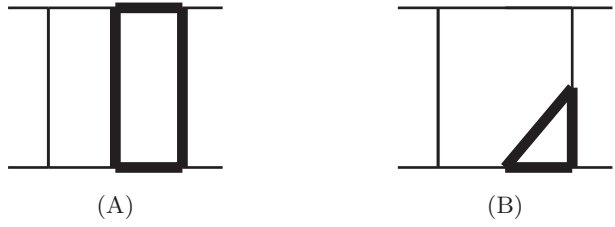
\caption{Seven-denominator topologies. Thin lines represent external massless particles and propagators, while thick lines
  represent massive propagators.
\label{fig1} }
\ec
\efig
%

Topologies A and B, do not complete the possible integrals involved in the planar QCD corrections to $q\bar{q} \to \gamma \gamma$.
Among them, there are three-point functions that are not included in the subtopologies of topologies A and B. These are the
three-point functions that occur for instance in the calculation of the NLO QCD corrections to the Higgs production in gluon fusion
(or Higgs decay into a pair of photons) and the corresponding MIs were studied in \cite{Aglietti:2006tp,Anastasiou:2006hc}.

\section{The System of Differential Equations \label{diff}}

For the analytic computation of the MIs, we use the differential equations method \cite{Kotikov:1990kg,Remiddi:1997ny,Gehrmann:1999as,Argeri:2007up,Henn:2014qga}. If we denote by  $\vec{f}(\vec{x},\epsilon)$ the vector of MIs found after the reduction process, we find that $\vec{f}(\vec{x},\epsilon)$
satisfies a system of first order linear differential equations with respect to the kinematic invariants $\vec{x}$, that we can write in differential form:
\be
d\vec{f}(\vec{x},\epsilon) = dA(\vec{x},\epsilon)\vec{f}(\vec{x},\epsilon) \, .
\label{de}
\ee
In Eq.~(\ref{de}), $A(\vec{x},\epsilon)$ is the matrix associated with the system,
which in general depends on the kinematic variables $\vec{x}$ and on the dimensional regulator $\epsilon$. The matrix $A(\vec{x},\epsilon)$ satisfies
the integrability conditions
\be
\partial_n A_m - \partial_m A_n - [A_n,A_m] = 0 \, ,
\ee
where $\partial_n = \frac{\partial}{\partial x_n}$, $A_n = \frac{\partial A}{\partial x_n}$ and $[A_n,A_m]=A_n A_m - A_m A_n$.

It was pointed out in \cite{Henn:2013pwa}, that in the case in which the master integrals can be expressed in terms of multiple polylogarithms,
the system can be cast into the following simplified form:
\be
 d\vec{f}(\vec{x},\epsilon) = \epsilon \, d\tilde{A}(\vec{x})\vec{f}(\vec{x},\epsilon) \, .
\ee 
$d\tilde{A}(\vec{x})$ is a d-log one-form, i.e. the entries of $\tilde{A}(\vec{x})$ are $\mathbb{Q}$-linear combinations of logarithms. In this basis the solution of the system is formally given in terms of Chen iterated integrals
\cite{Chen:1977oja}
\be
\vec{f}(\vec{x},\epsilon) = \mathbb{P}\exp\left(\epsilon\int_{\gamma}d\tilde{A}\right)\vec{f}_0(\epsilon) \, ,
\ee
where $\mathbb{P}$ stands for the path-ordered integration, $\gamma$ is some path in the kinematic invariants space and $\vec{f}_0(\epsilon)$ is a vector of boundary conditions.

We distinguish among two cases: the first one is the case in which the matrix $\tilde{A}(\vec{x})$ is rational in the kinematic invariants, while in the second case the dependence on the kinematic invariants can be algebraic, i.e. $\tilde{A}(\vec{x})$ depends also on roots of $\vec{x}$. 

In the first scenario, the entries are linear combinations of terms of the kind $\log( x_i-\alpha_k )$, where $\alpha_k$
are algebraic functions of kinematic invariants. In this case, once specified a path $\gamma:=\gamma(t)$, the solution 
can be written explicitly in terms of Goncharov's multiple polylogarithms \cite{Goncharov:polylog,Goncharov2007}
\be
\label{3.4}
G(\alpha_1,\cdots,\alpha_n;z) = \int^z_0 \frac{dt}{t-\alpha_1}G(\alpha_2,\cdots,\alpha_n;t) \, ,
\ee
with 
\be
\label{3.5}
G(\alpha_1;z) = \int^z_0\frac{dt}{t-\alpha_1} \;\;\;\; \alpha_1\neq 0 \, , \qquad G(\vec{0}_n;z) = \frac{\log^nz}{n!} \, .
\ee

In the second scenario, the entries contain roots of algebraic functions of kinematic invariants. In the case in which all the roots can be linearized simultaneously with a change of variables, the solution can be found, again, in terms of Goncharov's multiple polylogarithms.
If it is not possible to find such a transformation, one is left with repeated integrations over rational functions and/or multiple polylogarithms and a direct integration in terms of multiple polylogarithms is not easy. In this case, a possible strategy to arrive at the solution is to use the concept of symbol of an iterated integral in order to construct an ansatz in terms of multiple polylogarithms of a certain weight
(see for instance sect. 3.2 of \cite{Bonciani:2016qxi}).
Another strategy, is to follow the idea outlined in \cite{Caron-Huot:2014lda}. If the weight-2 is known in terms of Goncharov's multiple polylogarithms, we can integrate the following two integrations (up to weight 4) numerically. Moreover, integrating by-parts, we can reduce the double integration to a single integration, that involves irrational functions times weight-3 polylogarithms.
For the process under study, we found a transformation that linearizes a part of the roots present in the system of differential equations, but not all of them. In the set of new variables, the weight-2 is expressed in terms of multiple polilogarithms. It is therefore possible to follow the approach just outlined and to find the solution in terms of one- and two-fold numeric integrations.

\subsection{The Alphabet}

The matrices $\tilde{A}(\vec{x})$ of the systems of differential equations for the two topologies A and B depend on the following two groups of five square roots:
\bea
A & = & \left\{\sqrt{u\left(1+u\right)},\sqrt{u\left(-1+u\right)},\sqrt{v\left(v+1\right)},\sqrt{uv\left(uv + u + v\right)}, \right. \nonumber \\
&& \left. \sqrt{u (u+8 u v+16 (1+u) v^2)}\right\} \, , 
\label{Aset} \\
B & = & \left\{\sqrt{u\left(1+u\right)},\sqrt{u\left(-1+u\right)},\sqrt{v\left(v+1\right)},\sqrt{v\left(-1+v\right)}, \right. \nonumber \\ 
&& \left. \sqrt{uv\left(uv + u + v\right)}\right\} \, .
\label{Bset}
\eea

Regarding Topology B, we managed to further split system $B$ into two subsystems $B_1$ and $B_2$ that do not share master integrals, and have a smaller set of square roots:
\bea
\label{3.10a}
B_1 & = & \left\{\sqrt{u\left(u+1\right)},\sqrt{u\left(-1+u\right)},\sqrt{v\left(v+1\right)},\sqrt{uv\left(uv + u + v\right)}\right\} \, , \\
B_2 & = & \left\{\sqrt{u\left(u+1\right)},\sqrt{v\left(v+1\right)},\sqrt{v\left(-1+v\right)},\sqrt{uv\left(uv + u + v\right)}\right\} \, ,
\eea
which we could linearize simultaneously using the transformations (see Appendix \ref{appendixa})
\bea
u \; (\mbox{or} \; v) & \rightarrow & \frac{4 w^2 (w+1)^2}{\left(w^2+1\right) (w (5 w+4)+1)}
\label{3.10} \, , \\
v \; (\mbox{or} \; u) & \rightarrow & -\frac{16 w^4 (w+1)^2 z^2 (2 (w+1) z+w)^2}{\left(w^2+(w (w+2)-1) z\right) \left(w^2 (3 z-1)+2 w z+z\right) (w (3 w z+w+2 z)+z)} \times \nonumber \\
&& \times \frac{1}{(w (7 w z+w+6 z)+z)} \, .
\label{3.11}
\eea
For the subsystem $B_1$ the variable $u$ transforms according to \eqref{3.10} and the variable $v$ to the transformation \eqref{3.11}. Viceversa for the subsystem $B_2$. \par

For what concerns Topology A, it was not possible to split system $A$ into independent subsystems with a smaller number of roots.
We managed to linearize simultaneously the first four square roots, while it was not possible to linearize the fifth.
The presence of the fifth square root in Eq.~\eqref{Aset} can be found just in the differential equations that involve
the MIs of a 5-denominator four-point topology and the four-point topology at 7 denominators. In the rest of the system,
only the first four square roots appear.
This makes in such a way that we can separate the matrix of the system, $\tilde{A}(u,v)$ into two pieces 
\be
\tilde{A}(u,v) = \tilde{A}_{nr}(u,v) + \tilde{A}_r(u,v) \, . 
\ee 
The first term $\tilde{A}_{nr}(u,v)$ contains the same set of roots of the subsystem $B_1$, while $\tilde{A}_r(u,v)$ contains also the additional root $ \sqrt{u (u+8 u v+16 (1+u) v^2)}$. We performed the same change of variables used for the subsystem $B_1$. With this choice we could linearize the set of square roots of $\tilde{A}_{nr}(u,v)$ while the term $\tilde{A}_r(u,v)$ still contains square roots in the new kinematic variables $w$ and $z$.
This splitting of the system is functional to the solution. Indeed, the MIs whose differential equations are described by the entries of the matrix $\tilde{A}_{nr}(u,v)$ are analytically expressible in terms of GPLs, to all orders in the $\epsilon$ expansion. On the other hand, the MIs whose differential equations are described by the matrix $\tilde{A}_r(u,v)$ are not analytically expressible in terms of GPLs
and involve one- and two-fold numeric integrations.

For the GPL part of Topology A and for the entire Topology B, we have letters of the form $(w-w_k)$ and $(z-z_k)$, where
$w_k$ and $z_k$ belong to the following two sets, respectively:
\bea
\label{3.12}
w_k & :=& \biggl\{0,  -1,-\frac{2+i}{5},-\frac{2-i}{5},i,-i,1+\sqrt{2},1-\sqrt{2},- \frac{1+i \sqrt{2}}{3},-\frac{1-i \sqrt{2}}{3}  \biggr\} \, , \\
z_k &:=& \biggl\{0, -\frac{w}{2 (w+1)},-\frac{i w^2}{w^2-2 w-1},\frac{i w^2}{w^2-2 w-1},-\frac{w^2}{w^2+2 w-1},
-\frac{w^2}{3 w^2+2 w+1}, \nn\\
& & -\frac{i w^2}{(1+4 i) w^2-(2-4 i) w-1},\frac{w^2}{3 w^2+2 w+1},-\frac{w^2}{(4+i) w^2+(4-2 i) w-i},\frac{w^2}{A_w}, \nn\\
& & -\frac{w^2}{7 w^2+6 w+1}, \frac{-2 w^4-2 w^3-B_w}{9 w^4+12 w^3+10 w^2+4 w+1},\frac{-2 w^4-2 w^3+B_w}{9 w^4+12 w^3+10 w^2+4 w+1}, \nn\\
& & -\frac{w^2}{A_w},\frac{-4 w^4-4 w^3-C_w}{17 w^4+28 w^3+18 w^2+4 w+1},\frac{-4 w^4-4 w^3+C_w}{17 w^4+28 w^3+18 w^2+4 w+1},
\nn\\
& & \frac{2 w^4+2 w^3-iB_w}{w^4-4 w^3+2 w^2+4 w+1},\frac{2 w^4+2 w^3+iB_w}{w^4-4 w^3+2 w^2+4 w+1}\biggr\} \, ,
\eea
with
\bea
A_w &=& \sqrt{-w^4+4 w^3-2 w^2-4 w-1} \, , \\
B_w &=& \sqrt{-5 w^8-4 w^7-6 w^6-4 w^5-w^4} \, , \\
C_w &=& \sqrt{-w^8+4 w^7-2 w^6-4 w^5-w^4} \, .
\eea
The additional root $\sqrt{u (u+8 u v+16 (1+u) v^2)}$ in the new set of variables $w$ and $z$ becomes
\be
\label{3.13}
\sqrt{1 + \frac{E_w}{F_w}} \, ,
\ee
with
\bea
\label{3.14}
E_w & = & 128 w^4 (w+1)^2 z^2 (2 (w+1) z+w)^2 \left(w^8+8 (w+1) w^7 z+ \right. \nonumber \\
&& \left. +2 (w (5 w+6)+3) (w (7 w+2)+1) w^4 z^2+24 (w+1) (w (3 w+2)+1)^2 w^3 z^3+ \right. \nonumber \\
&& \left. + (w (3 w+2)+1)^2 (w (w (w (25 w+44)+26)+4)+1) z^4\right) \, , \\
F_w & = & \left(w^2+(w (w+2)-1) z\right)^2 \left(w^2 (3 z-1)+2 w z+z\right)^2 (w (3 w z+w+2 z)+z)^2 \times \nonumber \\
&& (w (7 w z+w+6 z)+z)^2 \, .
\eea

\section{The Master Integrals \label{mis}}

In order to find the canonical basis for the MIs several approaches exist \cite{Henn:2013pwa,Argeri:2014qva,Lee:2014ioa,Meyer:2016slj,Meyer:2017joq,Georgoudis:2016wff,Gituliar:2017vzm}.
In this paper we adopt the semi-algorithmic method described in \cite{Gehrmann:2014bfa}, which is based directly on the analysis
of the system of differential equations.
We put the system into canonical form using a ``bottom-up'' approach; we start from the subtologies with lowest non-trivial
number of denominators and we proceed with subtopologies with higher number of denominators, up to the completion of the rotation
of the whole system from pre-canonical to canonical form.
In general, we proceed following three steps:
\begin{itemize}

\item \emph{Step 1}. If we denote with $f_i(x)$ the MI under consideration, we choose the powers $a_i$ of the denominators in order to get the differential equation that concerns $f_i(x)$ in the following form
\be
\label{4.1}
df_i(x) = \left(H_{0,ij}(x) + \epsilon H_{1,ij}(x)\right)f_j(x) + \Omega_{ij}(x,\epsilon)g_{j}(x) \, .
\ee
The $\epsilon$ dependence of the non-homogeneous term $\Omega_{ij}(x,\epsilon)$ has to be of the form
\be
\label{4.2}
\Omega_{ij}(x,\epsilon) = \omega_{0,ij}(x) + \epsilon\omega_{1,ij}(x) +\sum_a \frac{\omega_{a,ij}(x)}{\epsilon+p_a} \, ,
\ee
where $p_a$ are real numbers.

\item{\emph{Step 2}. We remove the term $H_{0,ij}(x)$ rescaling the master $f_i(x)$ by a functions $h_{0,ij}(x)$ which satisfies the differential equation
\be
\label{4.3}
dh_{0,ij}(x) = -h_{0,ia}(x)H_{0,aj}(x) \, ,
\ee
where, again, the indices run over coupled master integrals. For all the masters that we studied we found that the functions $h_{0,ij}(x)$ are algebraic in the kinematic invariants. Therefore, the differential equation in the new master $\tilde{f}_i(x) = h_{0,ij}(x)f_{ij}(x)$ takes the form
\be
\label{4.4}
d\tilde{f}_i(x) = \epsilon \tilde{H}_{1,ij}(x)\tilde{f}_{j}(x)+\tilde{\Omega}_{ij}(x,\epsilon)g_j(x) \, .
\ee}

\item{\emph{Step 3}. We put in canonical form the non-homogeneous part shifting the MI $\tilde{f}_i(x)$ as
\be
\label{4.5}
\tilde{f}_i(x) \rightarrow \tilde{f}_i(x) + \left(\tilde{\omega}_{0,ij}(x) + \sum_a \frac{\tilde{\omega}_{a,ij}(x)}{\epsilon + p_a}\right)g_j(x) \, .
\ee
In order to remove the first and third term in \eqref{4.2}, the functions $\tilde{\omega}_{0,ij}(x)$ and $\tilde{\omega}_{a,ij}(x)$ must satisfy the following system of differential equations:
\bea
&& d\tilde{\omega}_{0,ij} - \tilde{H}_{1,ib}\tilde{\omega}_{a,bj} + \tilde{\omega}_{a,ib}G_{a,bj} + \omega_{0,ij} = 0 \, ,
\label{4.6} \\
&& d\tilde{\omega}_{a,ij} + p_a \tilde{H}_{ib}\tilde{\omega}_{a,bj} - p_a\tilde{\omega}_{a,ib}G_{1,bj} + \omega_{a,ij} = 0 \, ,
\label{4.7}
\eea
where $G_{1,ij}(x)$ is the matrix of the system of differential equations for the subtopology $g_j(x)$, which is already in caonical form:
\be
dg_{i}(x) = \epsilon G_{1,ij}(x)g_{j}(x) \, .
\label{4.8}
\ee

\bfig
\bc
\vspace*{-0.5cm}
\[ \vcenter{
\hbox{
  \begin{picture}(0,0)(0,0)
\SetScale{0.4}
  \SetWidth{1}
\CCirc(-15,15){5}{0.9}{0.9}
\CCirc(15,15){5}{0.9}{0.9}
  \SetWidth{4}
%
\CArc(-15,0)(15,0,180)
\CArc(-15,0)(15,180,360)
\CArc(15,0)(15,180,360)
%
\CArc(15,0)(15,0,180)
\Text(0,-22)[c]{\footnotesize{(${\mathcal T}_{1}$)}}
\end{picture}}
}
\hspace{1.4cm}
\vcenter{
\hbox{
  \begin{picture}(0,0)(0,0)
\SetScale{0.4}
  \SetWidth{1.0}
\DashLine(-35,0)(-20,0){3}
\DashLine(20,0)(35,0){3}
\CArc(0,0)(20,0,180)
\CArc(0,0)(20,180,360)
\CCirc(0,20){5}{0.9}{0.9}
\CCirc(34.5,23.5){5}{0.9}{0.9}
  \SetWidth{4}
\CArc(60,0)(40,150,180)
\CArc(0,34.6)(40,300,330)
\CArc(30.20,17.60)(5.28,-34,153)
\Text(-13,10)[c]{\footnotesize{$s,t$}}
\Text(0,-22)[c]{\footnotesize{(${\mathcal T}_2-{\mathcal T}_3$)}}
\end{picture}}
}
\hspace{1.55cm}
\vcenter{
\hbox{
  \begin{picture}(0,0)(0,0)
\SetScale{0.4}
  \SetWidth{1.0}
\DashLine(-35,0)(-20,0){3}
\DashLine(20,0)(35,0){3}
\CCirc(0,20){5}{0.9}{0.9}
\CCirc(34.5,23.5){5}{0.9}{0.9}
  \SetWidth{4}
\CArc(0,0)(20,0,180)
\CArc(0,0)(20,180,360)
\CArc(60,0)(40,150,180)
\CArc(0,34.6)(40,300,330)
\CArc(30.20,17.60)(5.28,-34,153)
\Text(-13,10)[c]{\footnotesize{$s$}}
\Text(0,-22)[c]{\footnotesize{(${\mathcal T}_4$)}}
\end{picture}}
}
\hspace{1.55cm}
\vcenter{
\hbox{
  \begin{picture}(0,0)(0,0)
\SetScale{0.4}
  \SetWidth{1.0}
\DashLine(-35,0)(-20,0){3}
\DashLine(20,0)(35,0){3}
\CCirc(0,20){5}{0.9}{0.9}
\CCirc(0,0){5}{0.9}{0.9}
\Line(-20,0)(20,0)
  \SetWidth{4}
\CArc(0,0)(20,0,180)
\CArc(0,0)(20,180,360)
\Text(-13,10)[c]{\footnotesize{$s,t$}}
\Text(0,-22)[c]{\footnotesize{(${\mathcal T}_5-{\mathcal T}_6$)}}
\end{picture}}
}
\hspace{1.55cm}
\vcenter{
\hbox{
  \begin{picture}(0,0)(0,0)
\SetScale{0.4}
  \SetWidth{1.0}
\DashLine(-35,0)(-20,0){3}
\DashLine(20,0)(35,0){3}
\CCirc(0,20){5}{0.9}{0.9}
\CCirc(0,-20){5}{0.9}{0.9}
\Line(-20,0)(20,0)
  \SetWidth{4}
\CArc(0,0)(20,0,180)
\CArc(0,0)(20,180,360)
\Text(-15,10)[c]{\footnotesize{$s,t$}}
\Text(0,-22)[c]{\footnotesize{(${\mathcal T}_{7}-{\mathcal T}_8$)}}
\end{picture}}
}
\hspace{1.55cm}
\vcenter{
\hbox{
  \begin{picture}(0,0)(0,0)
\SetScale{0.4}
  \SetWidth{1.0}
\DashLine(-45,0)(-30,0){3}
\DashLine(30,0)(45,0){3}
\CCirc(-15,15){5}{0.9}{0.9}
\CCirc(15,15){5}{0.9}{0.9}
\CArc(-15,0)(15,0,180)
\CArc(-15,0)(15,180,360)
  \SetWidth{4}
\CArc(15,0)(15,0,180)
\CArc(15,0)(15,180,360)
\Text(-14,10)[c]{\footnotesize{$s$}}
\Text(18,10)[c]{\footnotesize{$s$}}
\Text(0,-22)[c]{\footnotesize{(${\mathcal T}_{9}$)}}
\end{picture}}
}
\hspace{1.75cm}
\vcenter{
\hbox{
  \begin{picture}(0,0)(0,0)
\SetScale{0.4}
  \SetWidth{1}
\DashLine(-45,0)(-30,0){3}
\CCirc(-30,41){5}{0.9}{0.9}
\Line(15,30)(30,30)
\Line(15,-30)(30,-30)
  \SetWidth{4}
\Line(-30,0)(15,30)
\Line(15,-30)(-30,0)
\Line(15,30)(15,-30)
\CArc(-3.2,29)(40,185,227)
\CArc(-57.2,29)(40,313,355)
\CArc(-30,28)(13,-10,190)
\Text(-20,-6)[c]{\footnotesize{$s$}}
%
\Text(-2,-22)[c]{\footnotesize{(${\mathcal T}_{10}$)}}
\end{picture}}
}
\hspace{1.55cm}
\vcenter{
\hbox{
  \begin{picture}(0,0)(0,0)
\SetScale{0.4}
  \SetWidth{1}
\DashLine(-50,0)(-35,0){3}
\Line(10,30)(25,30)
\Line(10,-30)(25,-30)
\CCirc(-13,-15){5}{0.9}{0.9}
\CArc(-25,-34)(35,10,105)
\SetWidth{4}
\Line(-35,0)(10,30)
\Line(10,30)(10,-30)
\Line(10,-30)(-35,0)
\Text(-13,10)[c]{\footnotesize{$s,t$}}
\Text(-9,-13)[c]{\footnotesize{$3$}}
\Text(0,-28)[c]{\footnotesize{(${\mathcal T}_{11}-{\mathcal T}_{12}$)}}
\end{picture}}
}
%
\]
\vspace*{0.7cm}
\[
\vcenter{
\hbox{
  \begin{picture}(0,0)(0,0)
\SetScale{0.4}
  \SetWidth{1}
\DashLine(-50,0)(-35,0){3}
\Line(10,30)(25,30)
\Line(-35,0)(10,30)
\Line(10,-30)(25,-30)
\Line(10,-30)(-35,0)
\CCirc(10,0){5}{0.9}{0.9}
\SetWidth{4}
\CArc(30,0)(35,125,235)
\Line(10,30)(10,-30)
\Text(-13,8)[c]{\footnotesize{$s,t$}}
\Text(0,-28)[c]{\footnotesize{(${\mathcal T}_{13}-{\mathcal T}_{14}$)}}
\end{picture}}
}
\hspace{1.55cm}
\vcenter{
\hbox{
  \begin{picture}(0,0)(0,0)
\SetScale{0.4}
  \SetWidth{1}
\DashLine(-50,0)(-35,0){3}
\Line(10,30)(25,30)
\Line(-35,0)(10,30)
\Line(10,-30)(25,-30)
\Line(10,-30)(-35,0)
\CCirc(10,0){5}{0.9}{0.9}
\CCirc(-13,-15){5}{0.9}{0.9}
\SetWidth{4}
\CArc(30,0)(35,125,235)
\Line(10,30)(10,-30)
\Text(-13,8)[c]{\footnotesize{$s,t$}}
\Text(0,-28)[c]{\footnotesize{(${\mathcal T}_{15}-{\mathcal T}_{16}$)}}
\end{picture}}
}
\hspace{1.55cm}
\vcenter{
\hbox{
  \begin{picture}(0,0)(0,0)
\SetScale{0.4}
  \SetWidth{1.0}
\DashLine(50,0)(35,0){3}
\Line(-10,30)(-25,30)
\Line(-10,-30)(-25,-30)
\Line(-10,30)(-10,-30)
\Line(-10,-30)(35,0)
\CCirc(4,0){5}{0.9}{0.9}
%
  \SetWidth{4}
\CArc(-30,0)(35,305,55)
\Line(35,0)(-10,30)
\Line(35,0)(-10,-30)
\Text(18,8)[c]{\footnotesize{$s$}}
\Text(0,-28)[c]{\footnotesize{(${\mathcal T}_{17}$)}}
\end{picture}}
}
\hspace{1.55cm}
\vcenter{
\hbox{
  \begin{picture}(0,0)(0,0)
\SetScale{0.4}
  \SetWidth{1.0}
\DashLine(50,0)(35,0){3}
\Line(-10,30)(-25,30)
\Line(-10,-30)(-25,-30)
\Line(35,0)(-10,30)
\Line(35,0)(-10,-30)
\Line(-10,30)(-10,-30)
\Line(-10,-30)(35,0)
\CCirc(-10,0){5}{0.9}{0.9}
%
  \SetWidth{4}
\CArc(-30,0)(35,305,55)
\Line(35,0)(-10,30)
\Line(35,0)(-10,-30)
%
\Text(18,8)[c]{\footnotesize{$s$}}
\Text(0,-28)[c]{\footnotesize{(${\mathcal T}_{18}$)}}
\end{picture}}
}
\hspace{1.55cm}
\vcenter{
\hbox{
  \begin{picture}(0,0)(0,0)
\SetScale{0.4}
  \SetWidth{1.0}
\DashLine(50,0)(35,0){3}
\Line(-10,30)(-25,30)
\Line(-10,-30)(-25,-30)
\Line(-10,30)(-10,-30)
\Line(-10,-30)(35,0)
\CCirc(4,0){5}{0.9}{0.9}
\CCirc(18,10){5}{0.9}{0.9}
  \SetWidth{4}
\CArc(-30,0)(35,305,55)
\Line(35,0)(-10,30)
\Line(35,0)(-10,-30)
\Text(18,8)[c]{\footnotesize{$s$}}
\Text(0,-28)[c]{\footnotesize{(${\mathcal T}_{19}$)}}
\end{picture}}
}
\hspace{1.55cm}
\vcenter{
\hbox{
  \begin{picture}(0,0)(0,0)
\SetScale{0.4}
  \SetWidth{1.0}
\DashLine(-35,0)(-20,0){3}
\DashLine(20,0)(35,0){3}
\CArc(0,0)(20,0,360)
%
%
  \SetWidth{4}
\CArc(0,0)(20,270,360)
\CArc(0,0)(20,0,90)
\Line(0,-20)(0,20)
\Text(-13,10)[c]{\footnotesize{$s$}}
\Text(0,-28)[c]{\footnotesize{(${\mathcal T}_{20}$)}}
\end{picture}}
}
\hspace{1.55cm}
\vcenter{
\hbox{
  \begin{picture}(0,0)(0,0)
\SetScale{0.4}
  \SetWidth{1}
\DashLine(-45,0)(-30,0){3}
\Line(25,-30)(40,-30)
\Line(25,30)(40,30)
\CCirc(-15,15){5}{0.9}{0.9}
\CArc(-15,0)(15,0,180)
\CArc(-15,0)(15,180,360)
  \SetWidth{4}
\Line(0,0)(25,30)
\Line(25,-30)(0,0)
\Line(25,30)(25,-30)
\Text(-15,8)[c]{\footnotesize{$s$}}
\Text(0,-28)[c]{\footnotesize{(${\mathcal T}_{21}$)}}
\end{picture}}
}
\hspace{1.5cm}
\vcenter{
\hbox{
  \begin{picture}(0,0)(0,0)
\SetScale{0.4}
  \SetWidth{1.0}
\DashLine(-50,0)(-35,0){3}
\Line(10,30)(25,30)
\Line(-11,16)(-35,0)
\Line(10,-30)(25,-30)
\Line(-35,0)(10,-30)
%
  \SetWidth{4}
\Line(10,-30)(-11,16)
\Line(10,30)(-11,16)
\Line(10,-30)(10,30)
\Text(-15,8)[c]{\footnotesize{$s,t$}}
\Text(0,-28)[c]{\footnotesize{(${\mathcal T}_{22}-{\mathcal T}_{23}$)}}
\end{picture}}
}
\]
\vspace*{1.0cm}
\[
\vcenter{
\hbox{
  \begin{picture}(0,0)(0,0)
\SetScale{0.4}
  \SetWidth{1}
\Line(-25,30)(-40,30)
\Line(25,30)(-25,30)
\Line(-25,-30)(-40,-30)
\Line(-25,30)(-25,-30)
\Line(-25,-30)(25,-30)
\Line(25,-30)(25,-45)
\Line(25,30)(40,30)
\Line(25,-30)(25,30)
\CCirc(40,-10){5}{0.9}{0.9}
  \SetWidth{4}
\CArc(40,-25)(15,0,180)
\CArc(40,-25)(15,180,360)
%
%
%
\Text(0,-28)[c]{\footnotesize{(${\mathcal T}_{24}$)}}
\end{picture}}
}
\hspace{1.75cm}
\vcenter{
\hbox{
  \begin{picture}(0,0)(0,0)
\SetScale{0.4}
  \SetWidth{1}
\Line(-25,30)(-40,30)
\Line(-25,-30)(-40,-30)
\Line(25,-30)(40,-30)
\Line(25,30)(40,30)
\Line(-25,30)(-25,-30)
\Line(-25,-30)(25,-30)
\Line(25,30)(-25,30)
%
\CCirc(25,0){5}{0.9}{0.9}
  \SetWidth{4}
\CArc(45,0)(35,125,235)
\Line(25,-30)(25,30)
%
%
%
\Text(0,-28)[c]{\footnotesize{(${\mathcal T}_{25}$)}}
%
\end{picture}}
}
\hspace{1.7cm}
\vcenter{
\hbox{
  \begin{picture}(0,0)(0,0)
\SetScale{0.4}
  \SetWidth{1}
\Line(-25,30)(-40,30)
\Line(-25,-30)(-40,-30)
\Line(25,-30)(40,-30)
\Line(25,30)(40,30)
\Line(-25,30)(-25,-30)
\Line(-25,-30)(25,-30)
\Line(25,30)(-25,30)
\CCirc(25,0){5}{0.9}{0.9}
  \SetWidth{4}
\CArc(45,0)(35,125,235)
\Line(25,-30)(25,30)
%
%
%
\Text(0,-28)[c]{\footnotesize{(${\mathcal T}_{26}$)}}
\Text(0,20)[c]{\tiny{$(k_1\!-\!k_2)^2$}}
\end{picture}}
}
\hspace{1.7cm}
\vcenter{
\hbox{
  \begin{picture}(0,0)(0,0)
\SetScale{0.4}
  \SetWidth{1}
\Line(-25,30)(-40,30)
\Line(-25,-30)(-40,-30)
\Line(-25,-30)(25,-30)
\Line(25,-30)(40,-30)
\Line(25,30)(40,30)
\Line(-25,30)(-25,-30)
%
%
  \SetWidth{4}
\Line(-25,30)(25,-30)
\Line(25,-30)(25,30)
\Line(25,30)(-25,30)
%
%
%
\Text(0,-28)[c]{\footnotesize{(${\mathcal T}_{27}$)}}
%
\end{picture}}
} 
\hspace{1.7cm}
\vcenter{
\hbox{
  \begin{picture}(0,0)(0,0)
\SetScale{0.4}
  \SetWidth{1}
\Line(-25,30)(-40,30)
\Line(-25,-30)(-40,-30)
\Line(-25,-30)(25,-30)
\Line(25,-30)(40,-30)
\Line(25,30)(40,30)
\Line(-25,30)(-25,-30)
\CCirc(0,0){5}{0.9}{0.9}
  \SetWidth{4}
\Line(-25,30)(25,-30)
\Line(25,-30)(25,30)
\Line(25,30)(-25,30)
%
%
%
\Text(0,-28)[c]{\footnotesize{(${\mathcal T}_{28}$)}}
%
\end{picture}}
}
\hspace{1.55cm}
\vcenter{
\hbox{
  \begin{picture}(0,0)(0,0)
\SetScale{0.4}
  \SetWidth{1}
\Line(25,-30)(40,-30)
\Line(25,30)(40,30)
\Line(-25,30)(-40,30)
\Line(-25,-30)(-40,-30)
\Line(-25,30)(-25,-30)
\CCirc(-10,0){5}{0.9}{0.9}
  \SetWidth{4}
\CArc(-45,0)(35,-55,55)
\Line(25,-30)(25,30)
\Line(-25,-30)(25,-30)
\Line(25,30)(-25,30)
%
%
%
\Text(0,-28)[c]{\footnotesize{(${\mathcal T}_{29}$)}}
%
\end{picture}}
}
\hspace{1.55cm}
\vcenter{
\hbox{
  \begin{picture}(0,0)(0,0)
\SetScale{0.4}
  \SetWidth{1}
\Line(25,-30)(40,-30)
\Line(25,30)(40,30)
\Line(-25,30)(-40,30)
\Line(-25,-30)(-40,-30)
\Line(-25,30)(-25,-30)
\CCirc(-10,0){5}{0.9}{0.9}
  \SetWidth{4}
\CArc(-45,0)(35,-55,55)
\Line(25,-30)(25,30)
\Line(-25,-30)(25,-30)
\Line(25,30)(-25,30)
\Text(2,0)[c]{\footnotesize{$3$}}
%
\Text(0,-28)[c]{\footnotesize{(${\mathcal T}_{30}$)}}
\end{picture}}
}
\]
\vspace*{1.0cm}
\[
\vcenter{
\hbox{
  \begin{picture}(0,0)(0,0)
\SetScale{0.4}
  \SetWidth{1}
\Line(25,-30)(40,-30)
\Line(25,30)(40,30)
\Line(-25,30)(-40,30)
\Line(-25,-30)(-40,-30)
\Line(-25,30)(-25,-30)
\CCirc(-10,0){5}{0.9}{0.9}
  \SetWidth{4}
\CArc(-45,0)(35,-55,55)
\Line(25,-30)(25,30)
\Line(-25,-30)(25,-30)
\Line(25,30)(-25,30)
%
%
%
\Text(0,-28)[c]{\footnotesize{(${\mathcal T}_{31}$)}}
\Text(0,20)[c]{\tiny{$(k_1\!-\!k_2)^2$}}
\end{picture}}
}
\hspace{1.65cm}
\vcenter{
\hbox{
  \begin{picture}(0,0)(0,0)
\SetScale{0.4}
  \SetWidth{1.0}
\DashLine(-50,0)(-35,0){3}
\Line(-35,0)(10,-30)
\Line(-11,16)(-35,0)
\Line(-11,-16)(-35,0)
\Line(10,-30)(25,-30)
\Line(10,30)(25,30)
  \SetWidth{4}
\Line(-11,-16)(-11,16)
\Line(10,-30)(10,30)
\Line(10,30)(-11,16)
\Line(10,-30)(-11,-16)
\Text(-15,8)[c]{\footnotesize{$s$}}
\Text(0,-28)[c]{\footnotesize{(${\mathcal T}_{32}$)}}
\end{picture}}
}
\hspace{1.65cm}
%
\vcenter{
\hbox{
  \begin{picture}(0,0)(0,0)
\SetScale{0.4}
  \SetWidth{1}
\Line(-25,30)(-40,30)
\Line(-25,-30)(-40,-30)
\Line(-25,-30)(25,-30)
\Line(25,30)(-25,30)
\Line(25,-30)(40,-30)
\Line(25,30)(40,30)
\Line(-25,30)(-25,-30)
%
%
  \SetWidth{4}
\Line(25,-30)(0,30)
\Line(25,30)(0,30)
\Line(25,-30)(25,30)
%
%
%
\Text(0,-28)[c]{\footnotesize{(${\mathcal T}_{33}$)}}
%
\end{picture}}
}
\hspace{1.65cm}
%
%
%
\vcenter{
\hbox{
  \begin{picture}(0,0)(0,0)
\SetScale{0.4}
  \SetWidth{1}
\Line(25,-30)(40,-30)
\Line(25,30)(40,30)
\Line(-25,30)(-40,30)
\Line(-25,-30)(-40,-30)
\Line(-25,30)(-25,-30)
\Line(-25,-30)(25,-30)
\Line(25,30)(-25,30)
\Line(-25,-30)(-25,30)
%
%
  \SetWidth{4}
\Line(0,-30)(0,30)
\Line(0,-30)(25,-30)
\Line(25,-30)(25,30)
\Line(25,30)(0,30)
%
%
%
%
\Text(0,-28)[c]{\footnotesize{(${\mathcal T}_{34}$)}}
%
\end{picture}}
}
\hspace{1.65cm}
\vcenter{
\hbox{
  \begin{picture}(0,0)(0,0)
\SetScale{0.4}
  \SetWidth{1}
\Line(25,-30)(40,-30)
\Line(25,30)(40,30)
\Line(-25,30)(-40,30)
\Line(-25,-30)(-40,-30)
\Line(-25,30)(-25,-30)
\Line(-25,-30)(25,-30)
\Line(25,30)(-25,30)
\Line(-25,-30)(-25,30)
%
%
  \SetWidth{4}
\Line(0,-30)(0,30)
\Line(0,-30)(25,-30)
\Line(25,-30)(25,30)
\Line(25,30)(0,30)
\Text(0,20)[c]{\tiny{$(k_2\!+\!p_1)^2$}}
%
%
\Text(0,-28)[c]{\footnotesize{(${\mathcal T}_{35}$)}}
%
\end{picture}}
}
\hspace{1.65cm}
\vcenter{
\hbox{
  \begin{picture}(0,0)(0,0)
\SetScale{0.4}
  \SetWidth{1}
\Line(25,-30)(40,-30)
\Line(25,30)(40,30)
\Line(-25,30)(-40,30)
\Line(-25,-30)(-40,-30)
\Line(-25,30)(-25,-30)
\Line(-25,-30)(25,-30)
\Line(25,30)(-25,30)
\Line(-25,-30)(-25,30)
%
%
  \SetWidth{4}
\Line(0,-30)(0,30)
\Line(0,-30)(25,-30)
\Line(25,-30)(25,30)
\Line(25,30)(0,30)
\Text(0,20)[c]{\tiny{$(k_2\!+\!p_1)^2$}}
%
%
\Text(0,-28)[c]{\footnotesize{(${\mathcal T}_{36}$)}}
%
\end{picture}}
}
\hspace{1.75cm}
\vcenter{
\hbox{
  \begin{picture}(0,0)(0,0)
\SetScale{0.4}
  \SetWidth{1}
\Line(25,-30)(40,-30)
\Line(25,30)(40,30)
\Line(-25,30)(-40,30)
\Line(-25,-30)(-40,-30)
\Line(-25,30)(-25,-30)
\Line(-25,-30)(25,-30)
\Line(25,30)(-25,30)
\Line(-25,-30)(-25,30)
%
%
  \SetWidth{4}
\Line(0,-30)(0,30)
\Line(0,-30)(25,-30)
\Line(25,-30)(25,30)
\Line(25,30)(0,30)
\Text(0,20)[c]{\tiny{$(k_2\!+\!p_1)^2(k_2\!+\!p_1)^2$}}
%
%
\Text(0,-28)[c]{\footnotesize{(${\mathcal T}_{37}$)}}
%
\end{picture}}
}
\]
\vspace*{7mm}
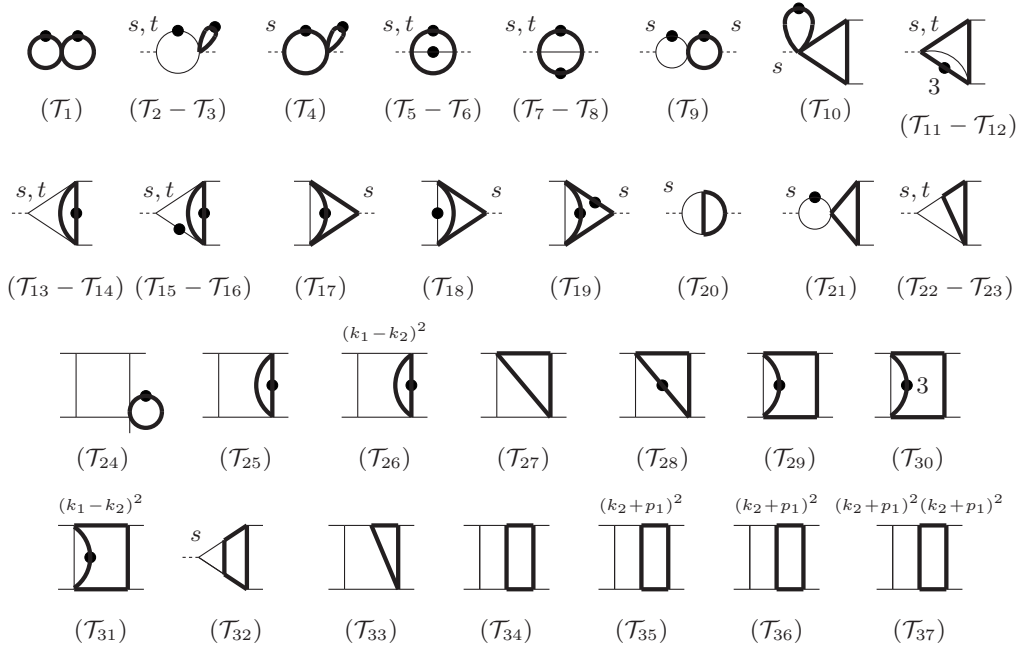
\caption{Master Integrals in pre-canonical form. Internal plain thin lines represent massless propagators,
  while thick lines represent the heavy-quark propagator. External plain thin lines represent massless particles
  on their mass-shell. Some of the masters, that depend on a single external variable, can appear both as function
  of $s$ or $t$. Because of this fact, we marked explicitly the functional dependence of the master with ``$s$'',
  ``$t$'' or ``$s,t$''.
\label{fig2}}
\ec
\efig
%
Although in principle the differential equations (\ref{4.6},\ref{4.7}) can be as difficult as the starting system for the MIs, in all the cases under study we managed to solve them and the solutions are algebraic functions of the kinematical invariants.}
\end{itemize}

Given the set of 37 MIs in pre-canonical form, as shown in Fig.~\ref{fig2}, we define the canonical basis, $f_1, ... , f_{37}$, from the following relations:
\bea
f_1 & = & \epsilon^2 {\mathcal T}_1 \, , \\
f_2 & = & \epsilon^2 s {\mathcal T}_2 \, , \\
f_3 & = & \epsilon^2 t {\mathcal T}_3 \, , \\
f_4 & = & \epsilon^2 \sqrt{s}\sqrt{s-4m_t^2} {\mathcal T}_4 \, , \\
f_5 & = & \epsilon^2 \sqrt{s}\sqrt{s-4m_t^2}({\mathcal T}_5 +\frac{1}{2}{\mathcal T}_7) \, , \\
f_6 & = & \epsilon^2 s {\mathcal T}_7 \, , \\
f_7 & = & \epsilon^2 \sqrt{t}\sqrt{t-4m_t^2}({\mathcal T}_6 +\frac{1}{2}{\mathcal T}_8) \, , \\
f_8 & = & \epsilon^2 t {\mathcal T}_8 \, , \\
f_9 & = & \epsilon^2 \sqrt{s}\sqrt{s(s-4m_t^2)} {\mathcal T}_9 \, , \\
f_{10} & = & -2s\epsilon^3 {\mathcal T}_{10} \, , \\
f_{11} & = & \epsilon^2 m_t^2 s {\mathcal T}_{11} \, , \\
f_{12} & = & \epsilon^2 m_t^2 t {\mathcal T}_{12} \, , \\
f_{13} & = & \epsilon^3 s {\mathcal T}_{13} \, , \\
f_{14} & = & \epsilon^2 \sqrt{s}\sqrt{s(s+4m_t^2)} {\mathcal T}_{15} - \epsilon^3 \frac{\sqrt{s}\sqrt{s+4m_t^2}}{2m_t^2(1+2\epsilon)}{\mathcal T}_1 \, , \\
f_{15} & = & \epsilon^3 t {\mathcal T}_{14} \, , \\
f_{16} & = & \epsilon^2 \sqrt{t}\sqrt{t(t+4m_t^2)} {\mathcal T}_{16} - \epsilon^3 \frac{\sqrt{t}\sqrt{t+4m_t^2}}{2m_t^2(1+2\epsilon)}{\mathcal T}_1 \, , \\
f_{17} & = & -\epsilon^3 \sqrt{s}\sqrt{s-4m_t^2} {\mathcal T}_{19} +\epsilon^2 m_t^2\sqrt{s}\sqrt{s-4m_t^2}{\mathcal T}_{17} -\frac{\epsilon^3}{2}\sqrt{s}\sqrt{s-4m_t^2}{\mathcal T}_{18} \, , \\
f_{18} & = & \epsilon^3 s {\mathcal T}_{17} \, , \\
f_{19} & = & \epsilon^3 s {\mathcal T}_{18} \, , \\
f_{20} & = & \epsilon^3 (1 - 2\epsilon) s {\mathcal T}_{20} \, , \\
f_{21} & = & \epsilon^3 s^2 {\mathcal T}_{21} \, , \\
f_{22} & = & \epsilon^4 s {\mathcal T}_{22} \, , \\
f_{23} & = & \epsilon^4 t {\mathcal T}_{23} \, , \\ 
f_{24} & = & \epsilon^3 st {\mathcal T}_{24} \, , \\
f_{25} & = & \epsilon^3 s\sqrt{t}\sqrt{t-4m_t^2}{\mathcal T}_{25} \, , \\
f_{26} & = & \epsilon^3 s {\mathcal T}_{26} -\frac{e^2}{2}s {\mathcal T}_8 \, , \\
f_{27} & = & \epsilon^4 (s+t) {\mathcal T}_{27} \, , \\
f_{28} & = & \epsilon^3 \sqrt{st}\sqrt{st-4m_t^2(s+t)}{\mathcal T}_{28} \, , \\
f_{29} & = & \epsilon^3 \sqrt{s}\sqrt{m_t^2 s-2 m_t^2 t (s+2 t)+s t^2}{\mathcal T}_{29} \, , \\
f_{30} & = & \epsilon^2 m_t^2 \sqrt{s} \sqrt{t} \sqrt{s t-4 m_t^2 (s+t)}{\mathcal T}_{30} -\epsilon^3 \sqrt{s} \sqrt{t} \sqrt{s t-4 m_t^2 (s+t)}{\mathcal T}_{29}  \, , \\
f_{31} & = & \epsilon^3 s {\mathcal T}_{31} \, , \\
f_{32} & = & \epsilon^4 s \sqrt{s(s-4m_t^2)}{\mathcal T}_{32} \, , \\
f_{33} & = & \epsilon^4 st {\mathcal T}_{33} \, , \\
f_{34} & = & \epsilon^4 s \sqrt{st}\sqrt{st - 4m_t^2(s+t)}{\mathcal T}_{34} + \epsilon^3 \sqrt{st} \sqrt{s t-4 m_t^2 (s+t)}{\mathcal T}_{28} + \epsilon^4 (s+t){\mathcal T}_{27}\\
& & + \epsilon^3 s {\mathcal T}_{31} + \epsilon^2 m_t^2 \sqrt{st} \sqrt{s t-4 m_t^2 (s+t)}{\mathcal T}_{30} - \epsilon^3 \sqrt{st} \sqrt{s t-4 m_t^2 (s+t)}{\mathcal T}_{29} \, , \\
f_{35} & = & \epsilon^4 s \sqrt{s(s-4m_t^2)} {\mathcal T}_{35} +\frac{\epsilon^3}{2}\sqrt{s}\sqrt{s-4m_t^2}{\mathcal T}_{28} + \epsilon^4 (s+t){\mathcal T}_{27} +\epsilon^3 s {\mathcal T}_{31} \\
& & + \epsilon^2 m_t^2 \sqrt{s} \sqrt{s-4 m_t^2} (s+2 t){\mathcal T}_{30} - \epsilon^3 \sqrt{s} \sqrt{s-4 m_t^2} (s+2 t) {\mathcal T}_{29} \, , \\
f_{36} & = & \epsilon^4 m_t^2 s^2 {\mathcal T}_{34} + \epsilon^4 s^2 {\mathcal T}_{36} +\epsilon^3 s {\mathcal T}_{31} + \epsilon^2 m_t^2 \sqrt{st} \sqrt{s t-4 m_t^2 (s+t)}{\mathcal T}_{30} \\
& & - \epsilon^3 \sqrt{st} \sqrt{s t-4 m_t^2 (s+t)} {\mathcal T}_{29}+ \epsilon^3 s^2 {\mathcal T}_{22} + \epsilon^3 s^2 {\mathcal T}_{21} \, , \\
f_{37} & = & \epsilon^4 s {\mathcal T}_{37} - \epsilon^4 st {\mathcal T}_{36} +\epsilon^4 (m_t^2 s -\frac{s}{2}){\mathcal T}_{35} - \epsilon^4 m_t^2 st {\mathcal T}_{34} + \epsilon^4 (s^2 +st){\mathcal T}_{32} \\
& & - \epsilon^3 \frac{s}{4}(s + 2t){\mathcal T}_{28} + \frac{\epsilon^4}{2}(-s + t){\mathcal T}_{27} + \epsilon^3 s {\mathcal T}_{31} - \frac{\epsilon^2}{2}m_t^2 s (s +2t){\mathcal T}_{30} \\
& & + \frac{\epsilon^3}{2}s(s + 2t){\mathcal T}_{29}+(1-2 \epsilon) \epsilon^3 \left(-\frac{s+t}{1-2 \epsilon}-s+t\right){\mathcal T}_{20} + \epsilon^3 s (s-t){\mathcal T}_{22} \\
& & + \epsilon^3 s (s-t){\mathcal T}_{21} + \epsilon^2 s \sqrt{s(s + 4m_t^2)}{\mathcal T}_{15} + \epsilon^3 (-s + t){\mathcal T}_{13} + 2 \epsilon^2 m_t^2 (s-t){\mathcal T}_{11} \\
& & + \epsilon^2 (-s + t){\mathcal T}_{7} +\epsilon^2 \left(\frac{\sqrt{s} \sqrt{4 m_t^2+s}}{4 (2 \epsilon+1) m_t^2}-\frac{\sqrt{s} \sqrt{4 m_t^2+s}}{4 m_t^2}\right){\mathcal T}_{1} \, .
\eea
The initial conditions $\vec{f}_0(\epsilon)$ for the MIs are fixed in the point $s=t=0$, where all the canonical master integrals vanish except for the double tadpole, which is known analytically, and masters $f_2$ and $f_3$ which are also known analytically, since they are product of two one-loop integrals.

The masters $f_1$--$f_{21}$ and $f_{24}$--$f_{26}$ were already present in the literature
\cite{Bonciani:2003te,Aglietti:2004tq,Bonciani:2003hc,Bonciani:2008ep,Bonciani:2007eh,Actis:2007fs}.
The masters $f_{29}$, $f_{30}$, $f_{31}$ were considered in \cite{Caron-Huot:2014lda}. The masters $f_{22}$, $f_{23}$, $f_{27}$, $f_{28}$, $f_{32}$, $f_{33}$, $f_{34}$, $f_{35}$,
$f_{36}$, $f_{37}$ to the best of our knowledge are new and presented in this article for the first time.
Among these, the subset $f_{22}$, $f_{23}$, $f_{27}$, $f_{28}$, $f_{32}$, $f_{33}$ can be written in terms of GPLs at every order in the dimensional parameter $\epsilon$. For the remaining ones, we were able to express the poles of
$f_{30}$--$f_{31}$ and $f_{34}$--$f_{37}$ in terms of GPLs, while the single pole of $f_{29}$ and the finite parts of
$f_{29}$--$f_{31}$ and $f_{34}$--$f_{37}$ still involve one- and two-fold numeric integrations.

The GPLs are found integrating the masters along the integration contour given by $\gamma = \gamma_w + \gamma_z$, where
\bea
\gamma_w (t) &:=& \left\{ \begin{array}{cc}
w(t) = tw & 0\leq t \leq1 \\
z(t) = 0 & \forall t
\end{array}\right\} \, , 
\label{4.53} \\
\gamma_z (t) &:=& \left\{ \begin{array}{cc}
w(t) = w & \forall t \\
z(t) = tz & 0\leq t \leq1
\end{array}\right\} \, .
\label{4.54}
\eea
The formal integration in $w$ and $z$ is made in the Euclidean region\footnote{This constraint comes from the linearization of the square roots in the differential equations system.}
\be
0 \leq w \leq -1+\sqrt{2}, \;\;\; 0 \leq z \leq \frac{w^2}{3 w^2+2 w+1} \, ,
\label{wzregion}
\ee
The formulas obtained, however, are valid in both the Euclidean and Minkowski regions, after analytical continuation, which can be done adding a small imaginary part to the Mandelstam variable $s$ according to the Feynman prescription $s+i0^+$.

In order to study the numerics of our analytic results in the Euclidean and Minkowski regions, we have to invert the relations (\ref{3.10},\ref{3.11})
finding $w$ as a function of $u$ and $z$ as a function of $v$, for fixed values of $u$. $w(u)$ and $z(u,v)$ have four different branches.
All the branches of $w(u)$ are discontinuous in the points
\be
\label{spoints}
u = \frac{4}{5}, \;\;\; u = 1 \, .
\ee
Therefore, in order to have the correct value of $w$ for $u$ spanning the complete domain, we must choose different branches of $w(u)$ and link them continuously in the singular points \eqref{spoints}. When $u \in \left[0,\frac{4}{5}\right]$, $w$ is given by
\bea
\label{wu1}
w(u) & = & -\frac{u-2}{5u-4} - \frac{u\sqrt{1+u}}{\sqrt{16-40u+25u^2}} + \frac{1}{2}\sqrt{A_u + B_u} \, ,
\eea
with
\bea
A_u & = & \frac{8 (u-2)^2}{(5 u-4)^2}-\frac{6 u}{5 u-4}-\frac{2 (3 u-2)}{5 u-4} \, ,  \\
B_u & = & -\frac{\sqrt{25 u^2-40 u+16} \left(-\frac{64 (u-2)^3}{(5 u-4)^3}+\frac{32 (3 u-2) (u-2)}{(5 u-4)^2}-\frac{32 u}{5 u-4}\right)}{8 u\sqrt{1+u}} \, .
\eea
and it varies in the real range $w \in \left[0,1\right]$. When $u \in \left[\frac{4}{5},\infty\right)$, $w$ is given by
\bea
\label{wu2}
w(u) & = & -\frac{u-2}{5u-4} + \frac{u\sqrt{1+u}}{\sqrt{16-40u+25u^2}} + \frac{1}{2}\sqrt{A_u + B_u} \, .
\eea
The domain $u \in \left[\frac{4}{5},1\right]$ is mapped into the real range $w \in \left[1, 1 + \sqrt{2}\right]$, while for $u > 1$ the variable $w$ becomes complex.
Regarding the variable $z$, we find that for any fixed value of $u$ we can map the whole Euclidean domain for $v$ into a closed finite range for $z$ using a single branch of $z(u,v)$.
The analytical continuation to the Minkowski region ($u<0$) is performed, according to the Feynman prescription, adding a small imaginary part to the Mandelstam variable $s$, $s+i0^+$. In particular, the range $u \in \left[0 , -1 \right]$ is mapped into a complex region for $w$, using the expression \eqref{wu1}.
Similarly, in the range $u \in \left[-1 , - \infty \right)$, using the expression \eqref{wu2}, we find that $u$ is mapped into a complex region of $w$.
  
We performed numerical checks of our GPL analytic results, comparing them with the results obtained using the
software FIESTA4 \cite{Smirnov:2008py,Smirnov:2013eza,Smirnov:2015mct} for points both in the Euclidean and Minkowski regions,
finding complete agreement. 

As mentioned previously, the appearance in the alphabet of the Topology A of the fifth root $\sqrt{u (u+8 u v+16 (1+u) v^2)}$ makes not possible to linearize simultaneously all the roots of the system. Therefore, we splitted the differential equation matrix into a sum of a part containing only GPL terms and a part that contains the root $\sqrt{u (u+8 u v+16 (1+u) v^2)}$.
The numerical integration is performed exploiting the fact that the weight 2 can be written in terms of GPLs for the whole system. This is because the fifth root affects the solution only from weight 3 on. Hence, we wrote the weights 3, $f_i^{(3)}(x)$ and 4, $f_{i}^{(4)}$, as single and double numerical integrations, respectively, over an analytic kernel of weight 2
\bea
\label{4.55}
f_{i}^{(3)}(x) &=& \int_{\gamma} dA_{ij}f_j^{(2)} \, , \\
\label{4.56}
f_{i}^{(4)}(x) &=& \int_{\gamma_1}dA_{ia}\int_{\gamma_2}dA_{aj}f_j^{(2)} \, .
\eea
For the numerical integrations, we chose the same paths that brought to the analytic expressions in terms of GPLs. We note that, in so doing, the integration in $w$ gives rise to expressions that are cast in polylogarithmic form. The sole numeric integration remains the one in $z$.

We checked our numerical results, coming from the numeric integrations, against the software FIESTA4 for points in the Euclidean region, finding complete agreement. The analytical continuation to the Minkowski region of the one- and two-fold parametric integrations is not considered in this article.

\section{Conclusions \label{concl}}

In this article we presented the calculation of the master integrals needed for the evaluation of the 
NNLO QCD planar corrections to di-photon (and di-jet) production in hadronic collisions. 

The system of differential equations satisfied by the masters is cast in canonical form and solved
in terms of Chen's iterated integrals. We represented these integrals in terms of Goncharov's multiple polylogarithms (GPLs), whenever it was possible to linearize the set of square roots appearing in the 
corresponding alphabet. However, the finite parts of seven four-point functions, and a single pole of one of them, still involve one- and two-fold numeric integrations of polylogarithms multiplied by irrational functions.
Choosing as a path for the iterated integration the same path that brought to the representation in terms of GPLs, we were able to integrate the letters that are function of the variable related to the partonic c.m. energy in closed form in terms of GPLs, remaining with an integration in the sole variable related to the partonic momentum transfert. 

The part expressed in terms of GPLs is integrated formally in a small part of the Euclidean region. However,
it can be evaluated in the whole Euclidean region and analitically continued to the Minkowskian region simply adding the correct causal prescription to the Mandelstam invariant $s$. We checked that the numerical values
obtained from our analytic expressions using the routines in \cite{Vollinga:2004sn} agree with the ones
obtaines with FIESTA4.

Concerning the numeric integrations, the fact that they involve only the variable related to the momentum transfert, allows for a strightforward evaluation in the whole Euclidean region and for a possibly
simple analytical continuation to the Minkowski region, that, nevertheless, we do not consider in the 
present work.

With the masters presented in this article (and other already present in the literature), it is possible to 
evaluate the leading color contribution of the massive corrections to the di-jet production. It is not 
possible to evaluate a gauge independent quantity in the di-photon production. For this we need the contribution of the crossed diagrams, that will be considered in a subsequent publication.

\section{Acknowledgments}

We would like to thank Francesco Moriello and Simone Lavacca for useful discussions about
the canonical form of the master integrals and the linearization of the square roots appearing
in the alphabet. Feynman diagrams are drawn with Axodraw \cite{Vermaseren:1994je}. 
A part of the necessary algebraic manipulations was preformed using FORM \cite{Vermaseren:2000nd}.

\appendix

\section{Linearization of Square Roots \label{appendixa}}

In order to write a solution in terms of multiple polylogarithms, the dependence of $\tilde{A}(\vec{x})$ on the kinematical invariants has to be rational.
Therefore, we need to find a change of variables that linearizes simultaneously all the roots in the matrices $\tilde{A}(\vec{x})$.
In this appendix, we describe the method that we used in order to obtain the transformations (\ref{3.10},\ref{3.11}), needed for the complete
linearization of square roots belonging to Topology B.

Instead of applying the procedure to the original set of roots \eqref{Bset}, we consider the following set
\be
\label{3.8a}
\left\{\sqrt{1+\tilde{u}},\sqrt{\tilde{u}-1},\sqrt{1+\tilde{v}},\sqrt{1+\tilde{u}+\tilde{v}}\right\} \, ,
\ee
where the kinematic variables $\tilde{u}$ and $\tilde{v}$ are the inverse of the variables $u$ and $v$, i.e.
\be
\label{3.8b}
\tilde{u} = -\frac{4 m_t^2}{s} = \frac{1}{u} \, , \;\;\; \tilde{v} = -\frac{4 m_t^2}{t}= \frac{1}{v} \, .
\ee
Once the transformations that linearize the set of roots \eqref{3.8a} are found, their inverse linearizes the set of roots \eqref{Bset}.

The problem of roots linearization is connected to the diophantine equation, which is a polynomial equation in many variables such that only its integers solutions are studied. It is possible to find a change of variables that linearizes simultaneously two or more roots solving iteratively
the diophantine equation associated to them. The idea is to transform the argument of the root into a perfect square, exploiting the solution of the diophantine equation associated as a parametrization for such transformation.

Consider the roots $\sqrt{1+\tilde{u}}$ and $\sqrt{\tilde{u}-1}$. The diophantine equation associated to the root $\sqrt{1+\tilde{u}}$ is
\be
1+\tilde{u}=p^2 \, ,
\label{3.8c}
\ee
where $p$ is an additional variable. Eq.~\eqref{3.8c} admits the integer solution $\tilde{u}=0$ and $p=-1$. The variable $p$ is promoted to a function $p(t_1,\tilde{u})$ of a parameter $t_1$, such that $p(t_1,0)=-1$. The easiest choice is the following linear parametrization:
\be
p=t_1 \tilde{u} - 1 \, .
\label{3.9} 
\ee
Substituting Eq.~\eqref{3.9} into Eq.~\eqref{3.8c}, we get a second order algebraic equation in the variable $\tilde{u}$ with $t_1$ as a parameter. The two solutions are 
\bea
\tilde{u} & = & 0 \, , \\
\tilde{u} & = & \frac{2 t_1+1}{t_1^2} \, .
\label{soldiop}
\eea
The second solution is exactly the transformation of $\tilde{u}$ that linearizes the root $\sqrt{1+\tilde{u}}$, as can be easily checked.

Once the first root is linearized, we consider the second root, $\sqrt{\tilde{u}-1}$. We transform the original variables according to the transformation that linearizes the first root, Eq.~(\ref{soldiop}). The diophantine equation associated to the root $\sqrt{\tilde{u}-1}$ is
\be
\tilde{u}-1=p^2
\ee
and, in the variable $t_1$, it becomes
\be
\frac{-t_1^2+2 t_1+1}{t_1^2} = p^2 \, .
\label{A3.12}
\ee
We note that the denominator of Eq.~\eqref{A3.12} is already a perfect square. Therefore, for our purpose it is sufficient to study the equation
\be
-t_1^2+2 t_1+1 = p^2 \, .
\label{A3.13}
\ee
We proceed as for the previous root, finding the integer solutions of \eqref{A3.13}, which are $t_1=0$ and $p=-1$ and parametrizing $p$ as
\be
p=t_2t_1-1 \, .
\label{A3.14}
\ee
Replacing Eq.~\eqref{A3.14} into Eq.~\eqref{A3.13}, we get again a second order algebraic equation with solutions
\bea
t_1 & = & 0 \, , \\
t_1 & = & \frac{2 (t_2+1)}{t_2^2+1} \, .
\label{soldiop2}
\eea
If we combine the two transformations
\be
\tilde{u} \rightarrow \frac{2 t_1+1}{t_1^2} \, , \quad  t_1 \rightarrow \frac{2 (t_2+1)}{t_2^2+1} \, ,
\ee
we find the transformation 
\be
\tilde{u} \rightarrow \frac{\left(t_2^2+1\right) (t_2 (t_2+4)+5)}{4 (t_2+1)^2} \, ,
\ee
that linearizes simultaneously the two roots $\sqrt{1+\tilde{u}}$ and $\sqrt{\tilde{u}-1}$.

For more than two square roots, we can repeat the procedure until all the roots are linearized.
Indeed, the restriction of the method is based on the existence of integer solutions to the diophantine equation.

The root $\sqrt{1+\tilde{v}}$ can be linearized with a transformation analogous to the second solution of Eq.~(\ref{soldiop}), with a different parameter $t$, $t_v$.

We consider now the last square root present in the set \eqref{3.8a},
$\sqrt{1+\tilde{u}+\tilde{v}}$. Assuming that the roots $\left\{\sqrt{\tilde{u}-1},\sqrt{1+\tilde{u}},\sqrt{1+\tilde{v}}\right\}$ are already linearized by means of the transformations
\bea
\tilde{u} & \rightarrow & \frac{\left(t_u^2+1\right) (t_u (t_u+4)+5)}{4 (t_u+1)^2} \, , \\
\label{3.22a}
\tilde{v} & \rightarrow & \frac{2 t_v+1}{t_v^2} \, ,
\label{3.22b}
\eea
the diophantine equation associated to the root $\sqrt{1+\tilde{u}+\tilde{v}}$ in the new variables is
\be
\left(t_u^2+1\right) (t_u (t_u+4)+5) t_v^2+4 (t_u+1)^2 t_v^2+4 (t_u+1)^2 (2 t_v+1)=p^2 \, .
\label{3.22c}
\ee
Equation \eqref{3.22c} is polynomial of degree 4 in the variable $t_u$ and of degree 2 in the variable $t_v$. Therefore, the solution of the diophantine
equation is more complicated with respect to the single variable case described in Eq.~\eqref{3.13}.
However, in this case we can find the following solution:
\be
t_v = 0, \;\;\; t_u = 0, \;\;\; p = -2 \, .
\label{3.22d}
\ee
Eq.~\eqref{3.22d} leads to the transformations \eqref{3.10} and \eqref{3.11}, once we consider the variables $u$ and $v$, that allow for a complete linearization of all the roots in the two subsystems of Topology B.

Let us consider now the roots appearing in the alphabet of Topology A. Beside the roots observed for Topology B, there is the additional root
$\sqrt{16\tilde{u} + (4 + \tilde{v})^2}$, which in the variables \eqref{3.8b} is $ \sqrt{u (u+8 u v+16 (1+u) v^2)}$. We note that this new root appears together with all the preceding square roots and it is not possible to split the set of coincident square roots in sub-sets that contain a smaller number of roots at the time.
Therefore we are forced to linearize $\sqrt{16\tilde{u} + (4+\tilde{v})^2}$ together with all the preceding four roots. However, the diophantine equation associated with the fifth root $\sqrt{16\tilde{u} + (4 + \tilde{v})^2}$ cannot be solved. The reason is that it is associated with a polynomial of degree 8 in the variable $z$ and degree 20 in the variable $w$, and it was not possible to find a parametrization, such as \eqref{3.22d}, which lower the degree of the polynomial equation so that it admits rational solutions.

Our alphabet in the new variables is constituted by linear letters and a single square root that remains in some of the
repeated integrations.

\section{Routing for the Pre-Canonical Master Integrals \label{appendixb}}

In this appendix we give the pre-canonical master integrals in the form \eqref{2.4}. The first list contains all the pre-canonical master integrals of the Topology A while the second list contains only the pre-canonical master integrals of the Topology B that are not present in the first list. As a consequence the denominators $D_i$ of the first list are relative to the Topology A while the denominators of the second list are relative to the Topology B.

\noindent Topology A:
\begin{align}
\mathcal{T}_1 & = \int \mathcal{D}^dk_1\mathcal{D}^dk_2 \frac{1}{D_4^2D_7^2} \, , 
& \mathcal{T}_2 & = \int \mathcal{D}^dk_1\mathcal{D}^dk_2 \frac{1}{D_2D_3^2D_7^2} \, , \\
\mathcal{T}_4 & =  \int \mathcal{D}^dk_1\mathcal{D}^dk_2 \frac{1}{D_4^2D_5D_6^2} \, ,
& \mathcal{T}_5 & =  \int \mathcal{D}^dk_1\mathcal{D}^dk_2 \frac{1}{D_3^2D_4D_5^2} \, ,\\
\mathcal{T}_6 & =  \int \mathcal{D}^dk_1\mathcal{D}^dk_2 \frac{1}{D_1^2D_4D_7^2} \, ,
& \mathcal{T}_7 & =  \int \mathcal{D}^dk_1\mathcal{D}^dk_2 \frac{1}{D_3D_4^2D_5^2} \, , \\
\mathcal{T}_8 & =  \int \mathcal{D}^dk_1\mathcal{D}^dk_2 \frac{1}{D_1D_4^2D_7^2} \, , 
& \mathcal{T}_9 & =  \int \mathcal{D}^dk_1\mathcal{D}^dk_2 \frac{1}{D_2D_3^2D_5^2D_6} \, , \\
\mathcal{T}_{10} & =  \int \mathcal{D}^dk_1\mathcal{D}^dk_2 \frac{1}{D_4^2D_5D_6D_7} \, , 
& \mathcal{T}_{11} & =  \int \mathcal{D}^dk_1\mathcal{D}^dk_2 \frac{1}{D_3D_4^3D_5D_7} \, , \\
\mathcal{T}_{12} & =  \int \mathcal{D}^dk_1\mathcal{D}^dk_2 \frac{1}{D_1D_4^3D_6D_7} \, , 
& \mathcal{T}_{13} & =  \int \mathcal{D}^dk_1\mathcal{D}^dk_2 \frac{1}{D_2D_3D_4^2D_7} \, , \\
\mathcal{T}_{15} & =  \int \mathcal{D}^dk_1\mathcal{D}^dk_2 \frac{1}{D_2D_3^2D_4D_7^2} \, , 
& \mathcal{T}_{17} & =  \int \mathcal{D}^dk_1\mathcal{D}^dk_2 \frac{1}{D_1D_4^2D_5D_6} \, , \\
\mathcal{T}_{18} & =  \int \mathcal{D}^dk_1\mathcal{D}^dk_2 \frac{1}{D_1^2D_4D_5D_6} \, , 
& \mathcal{T}_{19} & =  \int \mathcal{D}^dK_1\mathcal{D}^dk_2 \frac{1}{D_1D_4^2D_5D_6^2} \, , \\
\mathcal{T}_{20} & =  \int \mathcal{D}^dk_1\mathcal{D}^dk_2 \frac{1}{D_2D_3D_4D_5D_6} \, , 
& \mathcal{T}_{21} & =  \int \mathcal{D}^dk_1\mathcal{D}^dk_2 \frac{1}{D_2^2D_3D_5D_6D_7} \, , \\
\mathcal{T}_{22} & =  \int \mathcal{D}^dk_1\mathcal{D}^dk_2 \frac{1}{D_2D_3D_4D_6D_7} \, , 
& \mathcal{T}_{25} & =  \int \mathcal{D}^dk_1\mathcal{D}^dk_2 \frac{1}{D_1D_2D_3D_4^2D_7} \, , \\
\mathcal{T}_{26} & =  \int \mathcal{D}^dk_1\mathcal{D}^dk_2 \frac{D_8}{D_1D_2D_3D_4^2D_7} \, , 
& \mathcal{T}_{27} & =  \int \mathcal{D}^dk_1\mathcal{D}^dk_2 \frac{1}{D_1D_3D_4D_5D_7} \, , \\
\mathcal{T}_{28} & =  \int \mathcal{D}^dk_1\mathcal{D}^dk_2 \frac{1}{D_1D_3D_4^2D_5D_7} \, , 
& \mathcal{T}_{29} & =  \int \mathcal{D}^dk_1\mathcal{D}^dk_2 \frac{1}{D_1D_4^2D_5D_6D_7} \, , \\
\mathcal{T}_{30} & =  \int \mathcal{D}^dk_1\mathcal{D}^dk_2 \frac{1}{D_1D_4^3D_5D_6D_7} \, , 
& \mathcal{T}_{31} & =  \int \mathcal{D}^dk_1\mathcal{D}^dk_2 \frac{D_9}{D_1D_4^2D_5D_6D_7} \, , \\
\mathcal{T}_{32} & =  \int \mathcal{D}^dk_1\mathcal{D}^dk_2 \frac{1}{D_2D_3D_4D_5D_6D_7} \, , 
& \mathcal{T}_{34} & =  \int \mathcal{D}^dk_1\mathcal{D}^dk_2 \frac{1}{D_1D_2D_3D_4D_5D_6D_7} \, , \\
\mathcal{T}_{35} & =  \int \mathcal{D}^dk_1\mathcal{D}^dk_2 \frac{D_8}{D_1D_2D_3D_4D_5D_6D_7} \, , 
& \mathcal{T}_{36} & =  \int \mathcal{D}^dk_1\mathcal{D}^dk_2 \frac{D_9}{D_1D_2D_3D_4D_5D_6D_7} \, , \\
\mathcal{T}_{37} & =  \int \mathcal{D}^dk_1\mathcal{D}^dk_2 \frac{D_8D_9}{D_1D_2D_3D_4D_5D_6D_7} \, .
\end{align}

\noindent Topology B:
\begin{align}
\mathcal{T}_{3} & =  \int \mathcal{D}^dk_1\mathcal{D}^dk_2 \frac{1}{D_1D_6^2D_7^2} \, , 
& \mathcal{T}_{14} & =  \int \mathcal{D}^dk_1\mathcal{D}^dk_2 \frac{1}{D_1D_4^2D_6D_7} \, , \\
\mathcal{T}_{16} & = \int \mathcal{D}^dk_1\mathcal{D}^dk_2 \frac{1}{D_1D_4^2D_6D_7^2} \, , 
& \mathcal{T}_{23} & =  \int \mathcal{D}^dk_1\mathcal{D}^dk_2 \frac{1}{D_1D_4D_5D_6D_7} \, , \\
\mathcal{T}_{24} & =  \int \mathcal{D}^dk_1\mathcal{D}^dk_2 \frac{1}{D_1D_2D_3D_6^2D_7} \, , 
& \mathcal{T}_{33} & =  \int \mathcal{D}^dk_1\mathcal{D}^dk_2 \frac{1}{D_1D_2D_3D_4D_5D_6} \, .
\end{align}



\bibliographystyle{JHEP}
\bibliography{biblio}


\end{document}